\documentstyle[11pt,aaspp4]{article}

\def\AAA{\mbox{\AA}}

\slugcomment{To appear in May 1998 A.J.}

\lefthead{Gizis}
\righthead{Highly Active M subdwarfs}

\begin{document}

\title{Forever Young:  High Chromospheric Activity in M Subdwarfs\footnote
{Observations were made partially at the 60-inch telescope at Palomar
Mountain which is jointly owned by the California Institute of Technology
and the Carnegie Institution of Washington}
 }

\author{John E. Gizis\altaffilmark{2}}
\affil{Palomar Observatory, 105-24, California Institute of Technology,
  Pasadena, California 91125, e-mail: jeg@astro.caltech.edu}

\altaffiltext{2}{Current Address: LGRT 532A, Department of Physics and
Astronomy, University of Massachusetts, Amherst MA 01003-4525}
 
\begin{abstract}

We present spectroscopic observations of two halo M subdwarfs which
have $H \alpha$ emission lines.  We show that in both cases close companions
are the likely cause of the chromospheric activity in these old,
metal-poor stars.  We argue that Gl 781 A's unseen companion is most likely 
a cool helium white dwarf.  Gl 455 is a near-equal-mass M subdwarf (sdM) 
system.  
Gl 781 A is rapidly rotating with $v \sin i \approx 30$ km s$^{-1}$.   
The properties of the chromospheres and X-ray coronae of these systems
are compared to M dwarfs with emission (dMe).  The X-ray hardness ratios
and optical chromospheric line emission ratios are consistent 
with those seen in dMe stars.  Comparison to active near-solar metallicity
stars indicates that despite their low metallicity ($[m/H] \approx -1.2$),
the sdMe stars are roughly as active in both X-rays and chromospheric
emission.  Measured by $L_X/L_{bol}$, the activity level of 
Gl 781 A is no more than a factor of 2.5 subluminous with respect
to near-solar metallicity stars.  
\end{abstract}

\keywords{stars:activity --- stars: Population II --- stars:late-type ---
binaries:spectroscopic}

\section{Introduction \label{intro}}

It has long been noted that most old halo stars show little magnetic
activity relative to young disk stars --- indeed, Joy (1947)
used ``a weakening of emission lines'' as one of the criteria
for identifying M subdwarfs.  For disk stars, magnetic activity
has been extensively studied from the onset of the 
convective envelope amongst F stars down to 
the hydrogen burning limit (as reviewed by \cite{hs96}).
Comparatively little is known about the activity of Population II
stars --- recent studies of Population II stars 
have addressed chromospheric activity of single stars 
(\cite{ps97}) and X-ray emission in binary systems 
(\cite{ofp97}, hereafter OFP),
but these studies targeted near-solar mass stars rather than
very-low-mass stars.   Nevertheless, the Population II M subdwarfs
are worthy of study because stars with $M \lesssim 0.35 M_\odot$
are thought to be fully convective for all metallicities (\cite{cb97}),
and therefore may have a different dynamo mechanism than solar mass stars
which have radiative/convective zone 
boundaries (\cite{grkfsb96}; \cite{dyr93}).  

Moderate resolution spectra of very-low-mass stars allow 
the identification of  
high activity (``dMe'') stars by the presence of $H \alpha$ emission
lines while at the same time allowing the classification 
(\cite{g97}, hereafter G97) of metal-poor stars as sdM (M subdwarfs) 
and esdM (extreme M subdwarfs).
In our recent Palomar/Michigan State University (PMSU) Nearby-Star
Spectroscopic Survey (\cite{rhg95}; \cite{hgr96}), 
two of 2063 stars showed both $H \alpha$ emission
and anomalously weak TiO bands, indicating that these stars are sdMe.   
The PMSU sample consisted of the known main-sequence M stars within 25 parsecs,
and therefore was largely made up of disk near-solar metallicity stars.  
Subsequently, G97 surveyed a sample of cool high-velocity stars,
identifying an additional 27 sdM and 17 esdM.
Including our recent spectra of previously
unobserved faint high-proper-motion stars (\cite{gr97a}; \cite{gr97b}), 
we have observed a total of 68 sdM and esdM stars.  Only the
original two subdwarfs show emission.  G97 classified Gl 781 (LHS 482) 
as sdM1.5e and Gl 455 (LHS 2497) as sdM3.5e.  By fitting the synthetic spectra
of Allard and Hauschildt (1995), G97 showed that the sdM stars have
$[m/H] \approx -1.2 \pm 0.3$.  

We present additional observations of these two sdMe systems
aimed at understanding their nature and investigating the
characteristics of their activity.  Both stars prove to be close 
binary systems.  We present the data in 
Section~\ref{data}, discuss both the unusual Gl 781 system and 
the magnetic activity levels in Section~\ref{discussion},
and summarize our conclusions in Section~\ref{conclusions}.

\section{Data Analysis \label{data}}

Spectra were obtained using the echelle spectrograph (\cite{m85})
of the Palomar 60-in. telescope.  The red cross-dispersing
prisms were used, yielding wavelength coverage from $4700 \AAA$
to $9200\AAA$ with gaps between orders beyond $7000 \AAA$.
The 2 pixel resolution was $0.3 \AAA$.  
Gl 455 was observed on 31 December 1995, 4 January 1996, and twice on 
6 January 1996.   Gl 781 was observed 23 times 
on UT dates 08-09 August 1996; four
other observations were taken in July 1994.

The echelle data were reduced on a Sparc 5 running UNIX using
the ECHELLE FIGARO routines (\cite{tm98}).  
Radial velocities were found by cross-correlation with 
the M dwarf standards from Marcy \& Benitz (1989).  Comparison
of other M dwarfs observed on the same observing runs to 
high precision radial velocities 
show that the accuracy is $\pm 2$ km s$^{-1}$; these uncertainties will be
discussed in more detail in Gizis {\it et al.} (in prep.).  
The radial velocities are given in Table~\ref{table-radvel}.  

\begin{figure}
\plotone{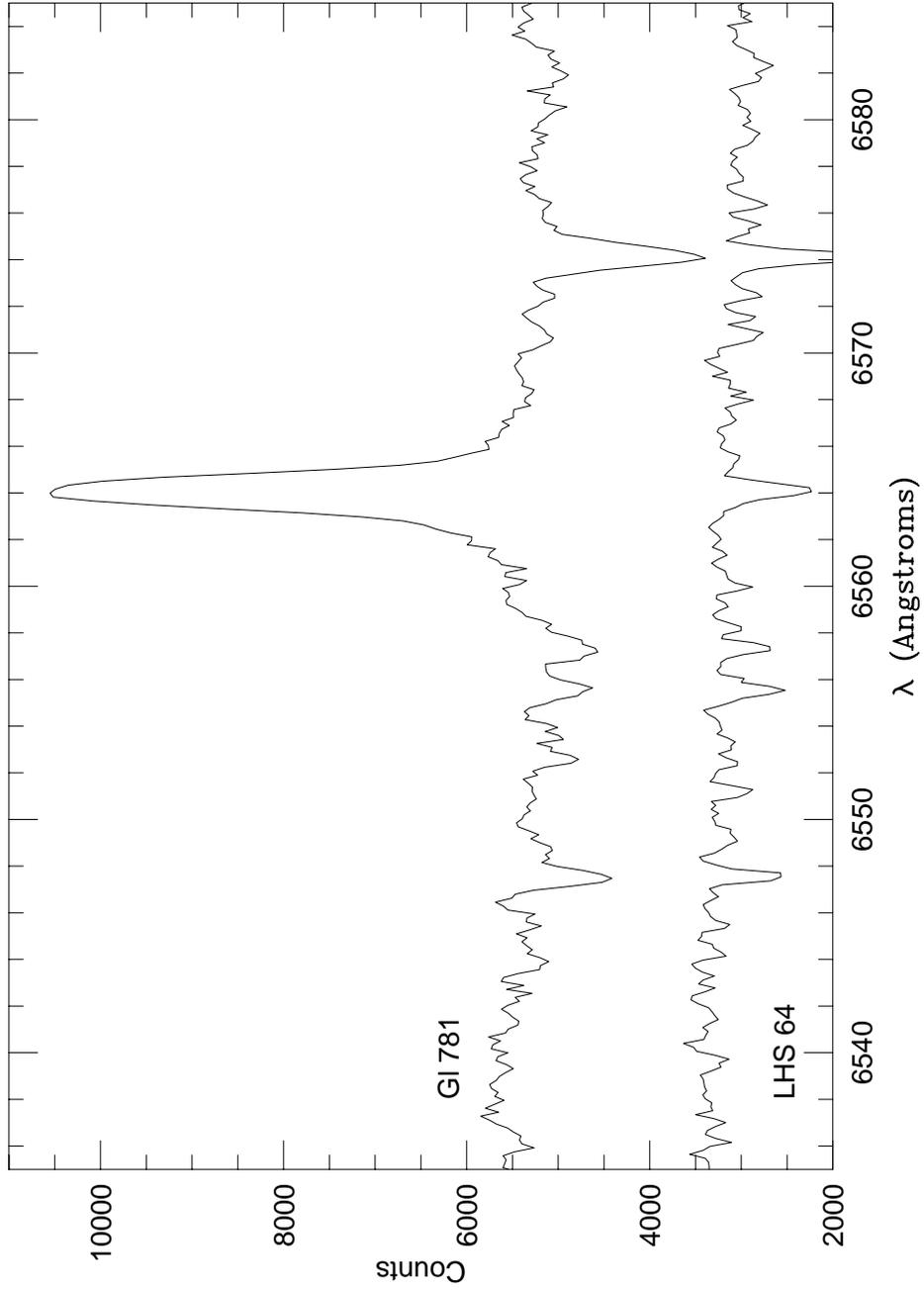}
\caption{Spectra of the sdM1.5 stars Gl 781 and LHS 64.  
LHS 64 has been shifted
by 338 km~s$^{-1}$ to match GL 781 A at this epoch.  Note the 
broadening of Gl 781 A's absorption lines with respect to LHS 64.
No lines from a companion to Gl 781 A are evident.     
\label{figure-gl781}}
\end{figure}

No lines from a companion were seen in the spectrum of Gl 781
(Figure~\ref{figure-gl781}), which
shows large velocity variations (\cite{j47}); however, 
our monitoring yields a reliable period estimate.  
T. Mazeh and D. Goldberg kindly calculated orbital solutions
using both June 1995 and August 1996 data.  Due to the large
separation in time between the observations, there are a number of 
different possible periods.  Orbital solutions for
each period are given in Table~\ref{orbital}.
Note that all periods lie within the range $0.497 \pm 0.001$ days.   
We will refer to the observed, brighter sdM star as Gl 781 A
and the unseen, fainter companion as Gl 781 B.   The mass of
Gl 781 A is likely to be $\sim 0.25 M_\odot$ according to the
models of D'Antona and Mazzitelli (1996) and Baraffe {\it et al.} (1995).  
With the observed mass function, $M \sin i = {0.3} M_\odot$
for Gl 781 B.  

The absorption lines in Gl 781 A are broadened compared to other
M dwarfs observed with the same setup.  
To measure this broadening, we artificially
broadened spectra of non-rotating stars obtained on the same night
by convolving with  the rotation profile (Gray 1992).  We used
the M1.5 V star Gl 880 as the standard.  Marcy and Chen (1992) 
determined the upper limit on this star's rotational velocity 
to be $1.1 \pm 1.1$ km s$^{-1}$.  
The least squares fit of the atomic lines yields 
$v \sin i = 30 \pm 5$ km s$^{-1}$.  To rule out the possibility of
metallicity differences causing the broadening, we have repeated this
procedure using the inactive sdM1.5 star LHS 64.   
LHS 64 does not show measurable
broadening relative to Gl 880, and fitting the Gl 781 spectra to 
artificially broadened LHS 64 spectra again 
yields $v \sin i = 30$ km s$^{-1}$.  We note, however, that 
Stauffer and Hartmann (1986) measured $v \sin i = 15$
km s$^{-1}$ for Gl 781 A.  Stauffer and Hartmann also measured
$v \sin i = 15$ km s$^{-1}$ for Gl 82 and Gl 268.  We also have
spectra of these stars using the same spectrograph setup ---
Gl 781 A's absorption lines are clearly broader.  We therefore
prefer our own measurement.  

The $H \alpha$ and $H \beta$ emission lines show significant variability.   
The $H \alpha$ line varies  
between $1.0 \AAA$ and $1.3 \AAA$ in the August 1996 spectra, 
with no dependence on orbital phase.
In the July 1994 spectra, $H \alpha$ is seen between $1.3$ and $1.6 \AAA$.   
Previous low resolution observations (\cite{rhg95}) observed $H \alpha$
in a stronger state ($1.9 \AAA$).  The $H \beta$ region is observed at low
signal to noise, but $H \beta$ is definitely also variable with 
equivalent width of $1.7 \AAA$ when $EW_{H \alpha} = 1.3$.  
Based on low-resolution spectrophotometry, obtained with the
Palomar 60-in. telescope and spectrograph in July 1996, we estimate the
continuum flux level as 
$1.1 \times 10^{-13}$ ergs cm$^{-2}$ s$^{-1}$ $\AAA^{-1}$ at $H \alpha$ 
and $4.7 \times 10^{-14}$ ergs cm$^{-2}$ s$^{-1}$ $\AAA^{-1}$ at
$H \beta$.  These indicate a line flux between 1.1 and 
$2.1  \times 10^{-13}$ ergs cm$^{-2}$ s$^{-1}$ for $H \alpha$,
and a Balmer decrement of 1.8 for the $EW_{H \alpha} = 1.3$ 
observation.
The He D3 line was detected with observed equivalent width of 
$0.15 \pm 0.5 \AAA$.  The He triplet at $6678 \AAA$ was also marginally
detected with an equivalent with of $\sim 0.1 \AAA$ (this line is 
difficult to measure as it lies near a TiO bandhead).  

Our four echelle spectra of Gl 455 clearly indicate it is 
a double-lined spectroscopic binary, but do not provide enough
data to determine the orbital parameters.
In the first two spectra (31 Dec. and 4 Jan.), the absorption lines
from both components 
are resolved, with the cores separated but the wings overlapping.
The other two spectra (6 Jan.) do not show line doubling.
The small shift ($0.9$ km s$^{-1}$) seen for the observations separated 
by two hours suggests the period is much longer than Gl 781's period. 
We note that the velocity data are consistent with a $\sim 8$ or
$\sim 16$ day period.
It should also be noted that since the lines are never completely separated we
cannot rule out the possibility that there is a (fainter) third  star 
in this system; G97 showed that the two components of Gl 455 appear
to lie above the sdM sequence in the HR diagram even after correcting
for binarity.  We do not detect rotational broadening in this system,
indicating that $v \sin i \le 15$ km s$^{-1}$.  Since our resolution is
insufficient to detect rotational broadening in	most dMe
(\cite{grh97}), this limit on the rotation is not surprising.  
If the period is 8 days, then we estimate that 
the equatorial rotational velocities are $\sim 1.3$ km s$^{-1}$.  

The $H \alpha$ emission of Gl 455 AB is not resolved, so
the reported equivalent widths are for the combined spectrum.  
In the December 1995
spectrum, the $H \beta$ line is resolved sufficiently to show that both 
components are in emission.  Significant variability is present:
the four equivalent widths of $H \alpha$ are $0.9, 1.1, 0.3, 0.6 \AAA$
and $H \beta$ are $2.0, 1.6, 0.3, 0.7 \AAA$.  Our earlier
moderate-resolution spectrum (\cite{hgr96}) had an $H \alpha$ equivalent
width of less than $1 \AAA$.  We therefore adopt the $H \alpha$
equivalent width of $0.6 \AAA$ as a ``typical'' value for the 
discussion of this system's activity level.  
Our July 1996 spectrophotometry
indicates that the continuum flux near $H \alpha$ for the two unresolved
components is approximately 
$5 \times 10^{-14}$ ergs cm$^{-2}$ s$^{-1}$ $\AAA^{-1}$.
We therefore estimate the ``typical'' $H \alpha$ flux to be  
$1.5 \times 10^{-14}$ ergs cm$^{-2}$ s$^{-1}$ for
each star under the assumption of equal-luminosity components.  

We have searched the ROSAT All-Sky Survey (\cite{1rxs}) to 
determine if the sdMe are X-ray sources.
We match Gl 781 with 1RXS J200503.8+542609 and
Gl 455 with 1RXS J120219.1+283507.  The hardness ratio (HR1) given
in the catalog is nearly identical with the hardness ratio (HR) 
used by Schmitt {\it et al.} 1995 to study nearby dM and dMe stars.
We therefore determine the count rate-to-energy 
conversion factor (CF) using their equation
$${\rm CF} = (5.30 {\rm HR1} + 8.31) \times 10^{-12} {\rm ergs~ cm^{-2}~
counts^{-1}}$$
The derived X-ray fluxes and other stellar parameters used in
this paper are given in Table~\ref{fluxes}.

\section{Discussion \label{discussion}}

\subsection{The Binary Gl 781 and Its History\label{discuss-history}}

The observed mass function of the system 
and the theoretically estimated mass of Gl 781 A imply that
Gl 781 B is more probably more massive 
($M_{B} \sin i \approx 0.3 M_\odot$).  Together
with the fact that no lines are seen, this mass 
implies that Gl 781 B is a faint,
cool white dwarf.  The line strengths, colors, and absolute magnitude of
Gl 781 A are similar to other sdM, so we estimate that 
the white dwarf is $\gtrsim 2$ magnitudes fainter, 
i.e. $M_V \gtrsim 13$. Ordinary white dwarfs ($M \sim 0.6 M_\odot$) 
have ages  $\gtrsim 10^9$ years at this luminosity 
(\cite{dm90}; we have used the
Liebert {\it et al.} 1988 bolometric corrections).  As shown
below, the most likely possibility is that Gl 781 B is a $\sim 0.35 M_\odot$
helium white dwarf --- the corresponding lower limit on the age is  
$2 \times 10^9$ years (\cite{hp97}).  

The detection of rotational broadening serves as a constraint on the
possible system parameters.  Assuming that the Gl 781 A's
rotational period is the same as that of the system,
$$\sin i = {{P \times (v_{rot} \sin i)}\over{2 \pi R_{\rm Gl 781 A}}}$$
The radius of the star is not known but can be estimated, either 
from stellar interior models, or by using the bolometric corrections
and effective temperatures from model atmospheres (\cite{ah95})
as fit by G97.  Using the latter method,
we estimate  $L \approx 4.6 \times 10^{31} {\rm erg~ s^{-1}}$
and $T_{eff} \approx 3600 \rm{K}$, so $R \approx 0.28 R_\odot$.
This estimate is in good agreement with stellar interior models
(\cite{bcah97}).  
We then find that we expect $v_{rot}$ to be $29 \pm 6$ km s$^{-1}$,
compared to the observed  $v_{rot} \sin i$ of $30 \pm 5$ km s$^{-1}$,
suggesting the system is viewed nearly edge on, with $\sin i \approx 1$.
If $\sin i$ is indeed near unity, then the white dwarf companion has an
unusually small mass.  Gravitational redshift measurements show that 
the distribution of white dwarf masses in wide binaries is 
strongly peaked at $M = 0.59 M_\odot$ (\cite{r96}). 
This peak mass cannot be reconciled with the radius of GL 781 A 
deduced from the observed luminosity and effective temperature.
Thus, the observations require that Gl 781 B is about half the mass of
the typical relatively isolated white dwarf.  

The small present day separation of the two stars, $a \approx 7 R_\odot$,
implies that Gl 781 A was engulfed when Gl 781 B became a red giant.
(Recall that the present day brighter component, the sdM,
was then the secondary by both mass and luminosity).  
There was therefore a phase of common envelope evolution.  
Since the mass of Gl 781 A is $\sim 0.25 M_\odot$ and 
the original mass of Gl 781 B was $M \gtrsim 1 M_\odot$ (since it 
has already evolved and cooled), it is clear that 
the mass ratio was greater than $\sim 4$
when the Gl 781 B became a red giant and first engulfed the companion.
Iben \& Livio (1993) have reviewed this situation and shown 
that a common envelope
will form and that the secondary (Gl 781 A) will not accrete a significant
amount of mass.  After being engulfed, Gl 781 A caused mass loss
from Gl 781 B, possibly ending B's evolution as a He white dwarf
with relatively low mass ($\sim 0.4 M_\odot$) instead of
allowing B to become a CO white dwarf ($M \approx 0.6 M_\odot$).  
This would be consistent with the estimate 
of $\sin i$ from the detected rotational broadening.   
The calculations of Iben {\it et al.} (1997) indicate that for
systems with 12 hour periods, helium white dwarfs with very-low-mass 
($M< 0.3M_\odot$) main sequence companions should be about seven times 
more common than CO white dwarfs with very-low-mass main sequence companions. 

\subsection{Chromospheric and Coronal Activity \label{discuss-activity}}

For M dwarfs, membership in a very close
($P \lesssim 5$ days) binary system is thought to be a sufficient 
condition for high chromospheric activity --- 
rapid rotation at relatively
old age is maintained by tidal interactions between the two stars
(\cite{b77}; \cite{ysh87}).  Both Gl 455 and Gl 781 prove to be short period
binaries, with Gl 781 A having detectable rotation, and we therefore 
attribute their activity to their binary nature.  
Gl 781 A's companion is not likely to affect the activity (except
by being the cause of Gl 781 A's rapid rotation).  
Although red dwarfs near hot white dwarfs show $H \alpha $ emission due
to reprocessing of the white dwarf's EUV flux
in the chromosphere of the red dwarf (\cite{szb96}),  
Gl 781 B's low luminosity 
($M_V>13$) implies that it is not a significant source of EUV flux.  
Furthermore,  Gl 781 A's 
emission is not stronger from the hemisphere facing the 
white dwarf, since there is no correlation between the strength of
$H \alpha$ with orbital phase in the August 1996 data.   
The white dwarf is therefore not causing the emission by 
incident radiation or other line-of-sight effects.
We also attribute the X-ray flux to a normal corona generated 
by Gl 781 A's rapid rotation, rather than accretion onto the
white dwarf, since at present
Gl 781 A is only about $10\%$ of the size of its Roche lobe
as estimated by Eggleton's (1983) formula.  We note also that
the X-ray flux is not due to some sort of ``basal'' emission 
unique to subdwarfs since
none of the known sdM and esdM without $H \alpha$ emission 
are detected in X rays by ROSAT.  

Metal-poor stars are usually associated with an extremely old 
($10^{10}$ yr) population.  Preston {\it et al.} (1996), however,
have identified young, A-type metal-poor main-sequence stars
with large velocities which they suggest are the residue
of a merger with a satellite galaxy (or galaxies).  
Presumably these stars have lower mass counterparts,
and if the M subdwarfs have emission lifetimes comparable to 
field M dwarfs (\cite{hgr96}), then these young M subdwarfs
will show detectable $H \alpha$ emission.   We cannot directly rule out
either Gl 455 or Gl 781 as members of this young, metal-poor population.  
The cooling time for the white dwarf in Gl 781 is at least 2 Gyr, 
which is too long to maintain the observed rapid rotation if M subdwarfs 
lose angular momentum like field M dwarfs.  Overall, the binary nature
of the two systems appears to be sufficient to explain their
activity.

The properties of sdMe relative to (near-solar metallicity) dMe  
are important because they provide a unique
perspective on the generation of strong activity.  Gl 781 A and
Gl 455 AB are the only known highly active, metal-poor, cool main
sequence stars.  Gl 781 A is of particular interest because its
high rotational velocity is similar to the rotational velocities
observed in the Pleiades cluster members of the same mass (\cite{jfs96}).
In comparing the properties of the sdMe, it should be be remembered that
considerable dispersion exists in the chromospheric and coronal
properties of even coeval, homogeneous samples such as the Pleiades
and Hyades clusters, and that even larger dispersion exists in 
field samples.  The physical causes of this dispersion are unknown.
Thus any relations based upon only two systems must
be viewed with caution.  

We first note that the data suggest that the chromospheres and 
coronae of the sdMe
are quite similar to disk dMe stars, despite the great metallicity
difference between the two classes.   The sdMe X-ray hardness ratios,
which are sensitive to the coronae temperatures, are
comparable to the field dMe with similar X-ray luminosities.  
The ratios of $H \alpha$ to $H \beta$ fluxes are similar to that seen in 
field dMe stars (\cite{hgr96}; \cite{grh97}).
The strength of the He D3 line is normal compared to the $H \alpha$ line,
as is the strength of the He 6678 line compared to the He D3 line.  
The correspondence between this sdMe and the dMe suggest that the
physical properties (e.g., temperature) of the 
lower chromosphere (traced by the Balmer lines), 
upper chromosphere (helium lines), and corona (X-ray)  
do not depend strongly on metallicity.    

We measure the overall activity of each star using the ratios of
the X-ray luminosity ($L_X$) and the $H \alpha$ luminosity 
($L_{H \alpha}$) to the star's bolometric luminosity.  
We focus on Gl 781 A since we know the orbital period 
and $v \sin i$, allowing a comparison to disk dMe stars with
similar rotation rates.  Gl 455 A and B are more slowly rotating
than Gl 781 A, but the period is unknown.  

Jones {\it et al.} (1996) have measured $v \sin i$ for 
M dwarfs in the Pleiades using the Keck I telescope.
The nine stars with $0.23 M_\odot \le M \le 0.29 M_\odot$,
similar to Gl 781 A's mass of $\sim 0.25 M_\odot$,
have measured velocities in the range 
8 km s$^{-1} \le v \sin i \le 41$ km s$^{-1}$.  
These results suggest that we can directly compare the 
metal-poor star Gl 781 A to a population of stars with similar masses and 
rotation rates.  Although this means two important parameters
are the same, there are three notable differences between 
Gl 781 A and the Pleiads:  first, Gl 781 A's age is most likely
10-15 Gyr, and certainly is 2 Gyrs or more which is $\sim 20$ times the 
age of the Pleiades cluster; second, Gl 781 A is metal-poor 
($[m/H] \approx -1.2$) whereas the Pleiades have near-solar metallicity;
third, Gl 781 A's companion might induce
internal and surface differential rotation different from that
of single stars (\cite{ysh87}).  

Direct comparison to the Pleiad sample observed by 
Jones {\it et al.} is hampered by the small sample size and
lack of X-ray data for most of their stars.  We instead compare to 
the much larger survey of the Pleiades cluster reported by
Hodgkin {\it et al.} (1995, hereafter HJS).  
The HJS survival analysis of both detections and upper limits
indicates that stars with masses near $0.25 M_\odot$
($M_I \approx 9$) have {\it mean} $\log L_X/L_{bol}$ 
and $\log L_{H\alpha}/L_{bol}$
of $\sim -3.3$ and $-3.7$.  The range in activity is large, as
some detected stars have $\log L_X/L_{bol} \approx -2$ while many stars
have upper limits indicating $\log L_X/L_{bol} < -3.5$.   
Gl 781 A's ratios are quite to close to the mean values, and on
this basis the activity of a rapidly-rotating metal-poor star appears to
be virtually identical to its more metal-rich counterparts --- 
although factors of $\sim 0.3$ in the log of the activity level
cannot be ruled out.
On the other hand, Gl 781 A is an order of magnitude
weaker in X-ray activity than the strongest M dwarfs in the Pleiades.
As noted by HJS, the reason why these sources are stronger 
is unknown.  Indeed, it interesting to note that many of
the slowest rotators in the Jones {\it et al.} sample have
the strongest $H \alpha$ emission --- an expanded data sample
would be of considerable interest to establish the nature of 
a rotation-activity relation for this stellar mass. We finally note that  
the Hyades cluster offers another point of comparison to 
the sdMe sample.  Again, the {\it mean} ratios of 
$L_X/L_{bol}$ and $L_{H\alpha}/L_{bol}$ for the Hyades
M dwarfs (\cite{rhm95}) are similar to the ratio seen
in Gl 781 A.  

We may also compare the activity of the sdM binaries to that of
nearby field dM binaries.  Like the M subdwarfs, these M dwarfs have 
activity induced by the rapid rotation caused by very close companions.
We have detected a further 16 SB2 M dwarf binaries using the same
echelle configuration as used for the current observations (\cite{grh97}).  
None of these systems have $v \sin i$ as great as Gl 781 A; all are likely
to have orbital periods less than 5-10 days.   
Matching these objects with the ROSAT All Sky Survey, we find that four have 
$\log (L_X/L_{bol}) < -3.3$ whereas twelve have $\log (L_X/L_{bol}) > -3.3$
(with four systems at the maximum of $\log (L_X/L_{bol}) \approx -2.9$).  
If rotation were the most important parameter, one might expect 
Gl 781 A to be as active in
X-rays as the most active nearby binaries.  In this case, Gl 781 A
is deficient by about 0.4 in the log (a factor of 2.5).  
It should also be recalled that the range in observed $H \alpha$ 
equivalent widths is 1.0 to 1.9, which suggests that some of 
the difference in activity level could be due to Gl 781 A
happening to be in a relatively weak state at the time of the ROSAT
observations.  Schmitt {\it et al.} (1995) have found that 
over a decade M dwarfs, unlike the Sun, do not show variability in X-rays in
excess of a factor of 2.  

One of the most striking aspects of the 
sdMe data is that both systems show similar X-ray activity levels
despite their different rotation rates.
Gl 455 AB certainly has a much slower rotation rate than Gl 781 A,
Indeed, $L_{H \alpha}/L_{bol}$ is much weaker in Gl 455 AB.  
The simplest explanation is that the Gl 455 AB X-ray observation is
affected by a flare or other transient event.  
There are a few M dwarfs in the Hyades (\cite{rhm95}) which have
$\log (L_{H \alpha}/L_{bol}) \approx -1$ for
non-simultaneous measurements, so Gl 455 AB's 
relative activity levels are not unprecedented.  

Comparing the sdMe to the near-solar mass Hyades binaries 
detected by Stern {\it et al.} (1995), 
Gl 781 A has an X-ray activity level
similar to that of the GK-dwarf ($B-V>0.6$) binaries
with periods of a few days but is less active than 
the half-day period binary V471Tauri.  Given the large mass
differences, we conclude merely that this example suggests that
it is not unreasonable to expect that Gl 781 A's half-day period
should put it amongst the most active binaries --- and therefore
that Gl 781 A may be mildly X-ray subluminous.  
There is, however, considerable evidence that the rotation-activity
relationship is complex in low-mass stars (indeed, Hawley {\it et al.}
1996 questioned the existence of a correlation between 
rotation rate and activity level in M dwarfs).  Stauffer {\it et al.} (1997)
have found that for a sample of Hyades M dwarfs ($M \approx 0.45 M_\odot$;
above the fully-convective limit and more massive than the sdMe)
that there is no single activity-rotation relation.  They propose
that their data can be explained as a ``bifurcation'' with 
two different rotation-activity relations.  On one relation,
relatively slowly rotating stars ($v \sin i < 6$ km s$^{-1}$) 
can be as active ($\log L_{H \alpha}/L_{bol} > -4.15$) as
stars (on the other relation)
with $10 \lesssim v \sin i \lesssim 15$ km s$^{-1}$.  
They suggest that the difference is due to differing rotational 
evolution histories which results in 
differing radial and latitudinal differential rotation.
They also find that stars with companions
as distant as 1 A.U. have enhanced activity relative to 
single stars with the same rotation rate.   In addition, there
is evidence that the most rapid ($P \lesssim 10$ hr.) 
GK rotators may weaken in activity level
(\cite{rsps96}).  Clearly, there are many 
parameters relevant in determining the activity level --- these 
or others may be important for Gl 781 A.  

We conclude that Gl 781 A does not show evidence for a significant
(i.e., factor of 10) weakening in its activity levels 
relative to near-solar metallicity stars.  It may, however, be
subluminous in X-rays for its bolometric luminosity by up 
to a factor of $\sim 2.5$.  This may be the result of its metal deficiency.  
Definite conclusions are prevented by the large, and as yet 
poorly-understood, dispersion in the activity levels of M dwarfs.  
The relatively small effect on the X-ray emission due to 
a factor of $\gtrsim 10$ deficiency in metals suggests that 
composition is not a major contributor to the dispersion in
activity levels observed for M dwarfs, which are all relatively 
close to solar metallicity ($[m/H] \gtrsim -0.6$, \cite{g97}). 

The OFP survey of spectral type FGK metal-poor binaries has 
shown that a sample of Population II 
short-period binaries analogous to RS CVn binaries 
are at least one order of magnitude less X-ray 
luminous than the disk RS CVn binaries, although some Population II 
binaries are as luminous as the RS CVn binaries.  They have suggested
that the X-ray line emission is reduced due to the relative lack of metals 
in the corona of Population II systems with $-1.4 < [m/H] < -0.4$.
Our sdMe lie in the low-metallicity end of that range. 
Based upon our discussion above, we conclude that the sdMe are
not as X-ray subluminous as expected by analogy with the OFP results.  

\section{Conclusions \label{conclusions}}

We have obtained spectra of two halo very-low-mass metal-poor systems 
that show $H \alpha$ and X-ray emission.  
Both prove to be close binary systems, which accounts for their 
high activity level.  
We deduce that Gl 781 A's unseen companion is a cool white dwarf.
The present-day orbital period of 12 hours implies that the
system passed through a phase of common envelope evolution.  
Gl 455 AB consists of two near-equal luminosity sdMe.  

The strength of the X-ray coronae of the sdMe relative to the
$H \alpha$ chromosphere is well within the range observed in
dMe stars in the field and in young clusters.  Gl 781 A's level of activity,
as measured by the ratio $L_X/L_{bol}$, is comparable to the mean
activity level in the Pleiades.   The comparison to both the Pleiades
and field dMe binaries suggest that Gl 781 A is no more than a factor of
$\sim 2.5$ subluminous in X-rays due to its metallicity.  This is contrast 
to the factor of ten observed in most (but not all) stars in the OFP 
survey of more massive metal-poor binaries.   

With only two sdMe systems, these results are necessarily uncertain.
Unfortunately, it will be difficult to significantly increase the
sample of sdMe and esdMe since the two systems discussed here
are the only such binaries detected amongst a total of 68
LHS catalog sdM and esdM stars observed from Palomar 
(\cite{g97}; \cite{gr97a}; \cite{gr97b}).   A more optimistic view,
however, is that activity due to rapid rotation could
allow the identification of, or place limits upon the existence of,  
short-period binaries and/or relatively young (few Gigayear) 
halo populations given a large spectroscopic sample.  

\acknowledgments

I would like to thank the staff of Palomar Observatory
and in particular Skip Staples for assistance with the
observations.  Jim McCarthy gave valuable tutorials on his
echelle data reduction software.  
T. Mazeh and D. Goldberg kindly computed the orbital 
solutions.  Neill Reid and Suzanne Hawley gave helpful comments.  
Mark Giampapa's referee review substantially improved the paper.  
I gratefully acknowledge support by Greenstein and
Kingsley Fellowships as well as NASA grants GO-06344.01-95A and
GO-05913.01-94A.
This research has made use of the Simbad database, operated at
CDS, Strasbourg, France.




\begin{table}
\dummytable\label{table-radvel}
\end{table}

\begin{deluxetable}{rrrr}
\tablewidth{0pc}
\tablenum{1A}
\tablecaption{Velocity Data for Gl 781}
\tablehead{
\colhead{JD} &
\colhead{$V_{rad}$} &
\colhead{JD} &
\colhead{$V_{rad}$} 
}
\startdata
 2449559.787284 & 70.59  & 2450304.002925 & -134.41 \nl 
 2449560.793864 & 58.07  & 2450304.022841 & -150.44 \nl 
 2449561.789240 & 57.89  & 2450304.661833 & -47.03 \nl 
 2449561.856991 & -37.40 & 2450304.707263 & 20.89 \nl 
 2450303.661066 & -61.09 & 2450304.757906 & 75.85 \nl 
 2450303.685817 & -23.23 & 2450304.801232 & 87.79 \nl 
 2450303.710209 & 16.40  & 2450304.806716 & 86.17 \nl 
 2450303.715577 & 22.64  & 2450304.811984 & 85.73 \nl 
 2450303.742898 & 58.07  & 2450304.816111 & 82.87 \nl 
 2450303.773024 & 77.66  & 2450304.872800 & 34.81 \nl 
 2450303.821616 & 84.31  & 2450304.925684 & -39.83 \nl 
 2450303.881348 & 32.56  & 2450304.975199 & -111.42 \nl 
 2450303.932413 & -42.01 & 2450305.021252 & -152.47 \nl 
 2450303.973081 & -102.64 \nl 
\enddata
\end{deluxetable}

\begin{deluxetable}{rrrrc}
\tablewidth{0pc}
\tablenum{1B}
\tablecaption{Velocity Data for Gl 455}
\tablehead{
\colhead{JD} &
\colhead{$V_{rad}^A$} &
\colhead{$V_{rad}^B$} &
\colhead{$\Delta V_{rad}$} & 
\colhead{Comment} 
}
\startdata
 2450083.053575 & 52.9 & -1.6 & 54.5 & Resolved \nl 
 2450087.002654 & -1.1 & 57.7  & 58.7 & Resolved\tablenotemark{a} \nl 
 2450088.909443 & 29.2 & \nodata & $<15$ & Unresolved \nl 
 2450088.996125 & 30.1 & \nodata & $<15$ & Unresolved  \nl 
\enddata
\tablenotetext{a}{The spectra of the two components are sufficiently similar
that the identification of ``A'' and ``B'' may be incorrect.}
\end{deluxetable}

\begin{deluxetable}{crrrrrrrcrrrrrlr}
\tablewidth{0pc}
\tablenum{2}
\tablecaption{Orbital Solutions for Gl 781}
\label{orbital}
\tablehead{
\colhead{ } &
\colhead{P} &
\colhead{$\gamma$} &
\colhead{K} & 
\colhead{e} & 
\colhead{$\omega$} & 
\colhead{$T_0$} & 
\colhead{$f(m)$}
}
\startdata
         A &  0.49603807 &  -33.97 & 123.31 &  0.0136 &  212. & 2450193.974 & 0.0965 \nl
 $\sigma_A$       &  0.00000057 &    0.40 &   0.63 &  0.0043 &   17. &    0.024 & 0.0015 \nl
         B &  0.49636949 &  -34.06 & 123.31 &  0.0123 &  212. &  2450194.397 & 0.0966 \nl
 $\sigma_B$    &  0.00000050 &    0.35 &   0.56 &  0.0038 &   17. &    0.024 & 0.0013 \nl
         C &  0.49670134 &  -34.15 & 123.29 &  0.0109 &  212. &  2450194.323 & 0.0966 \nl
 $\sigma_C$       &  0.00000049 &    0.34 &   0.54 &  0.0037 &   19. &    0.026 & 0.0013 \nl
         D &  0.49703363 &  -34.23 & 123.27 &  0.0095 &  211. &  2450194.249 & 0.0966 \nl
 $\sigma_D$       &  0.00000052 &    0.36 &   0.57 &  0.0039 &   23. &    0.031 & 0.0013 \nl
         E &  0.49736636 &  -34.31 & 123.24 &  0.0081 &  210. &  2450194.174 & 0.0966 \nl
 $\sigma_E$       &  0.00000059 &    0.42 &   0.66 &  0.0045 &   31. &    0.043 & 0.0015 \nl
         F &  0.49769953 &  -34.38 & 123.20 &  0.0066 &  208. &  2450194.097 & 0.0966 \nl
 $\sigma_F$       &  0.00000070 &    0.49 &   0.77 &  0.0052 &   45. &    0.062 & 0.0018 \nl
         G &  0.49803314 &  -34.45 & 123.16 &  0.0051 &  204. &  2450194.018 & 0.0966 \nl
 $\sigma_G$       &  0.00000082 &    0.58 &   0.91 &  0.0061 &   70. &    0.097 & 0.0021 \nl
\enddata
\end{deluxetable}

\begin{deluxetable}{cccc}
\tablewidth{0pc}
\tablenum{3}
\tablecaption{Stellar Parameters}
\label{fluxes}
\tablehead{
\colhead{Data} &
\colhead{units} &
\colhead{Gl 781 A} &
\colhead{Gl 455 AB} 
}
\startdata
V                                    & & 11.98 & 12.85\tablenotemark{a} \nl
V-I$_C$                              & & 1.99  & 2.47\tablenotemark{a} \nl
d                                 & pc & 16.6  & 20.3 \nl 
$M_V$                                & & 10.88 &12.06\tablenotemark{b} \nl
$M_{\rm bol}$                        & & 9.58  &  10.50\tablenotemark{b} \nl
Luminosity   & $10^{31}$ ergs s$^{-1}$ & 4.6   &  2.0\tablenotemark{b} \nl
$T_{eff}$                          & K & 3600  & 3500\tablenotemark{b} \nl
Radius                   & R$_{\odot}$ & 0.28  & 0.20\tablenotemark{b} \nl 
Mass\tablenotemark{c}      & $M_\odot$ & 0.25  & 0.15\tablenotemark{b}\nl
HR1                                  & & -0.41 & -0.74\tablenotemark{a} \nl
$f_X$ & $10^{-13}$ ergs cm$^{-2}$ s$^{-1}$ & 6.60 & 3.91\tablenotemark{a} \nl 
$L_X$   & $10^{27}$ ergs s$^{-1}$ & 22   &  9.7\tablenotemark{b} \nl
$\langle EW_{H\alpha}\rangle$\tablenotemark{d} & $\AAA$ & 1.3   & 0.6\tablenotemark{a} \nl
{$\log(L_{H\alpha}/L_X)$}            & & -0.6 & -1.1 \nl
{$\log(L_{H\alpha}/L_{\rm bol})$}    & & -3.9  & -4.4 \nl
{$\log(L_{X}/L_{\rm bol})$}          & & -3.3  & -3.3 \nl
$v \sin i$               & km s$^{-1}$ & 30    & $\le 15$ \nl
$P_{orb}$                     & {days} & 0.5   & 8? \nl
\enddata
\tablenotetext{a}{Value for unresolved system}
\tablenotetext{b}{Value for individual star A or B
 assuming equal-luminosity components}
\tablenotetext{c}{Estimated from stellar models}
\tablenotetext{d}{``Typical'' value used to estimate activity levels involving
$L_{H\alpha}$}
\end{deluxetable}

\clearpage

\end{document}